\documentclass[prl,twocolumn,superscriptaddress,amsmath,amssymb,longbibliography]{revtex4-2}

\usepackage{graphicx}
\usepackage{xcolor}
\usepackage{bm}
\usepackage{mathtools}
\usepackage{amsthm}
\newtheorem{theorem}{Theorem}
\newtheorem{lemma}{Lemma}
\usepackage[colorlinks=true,linkcolor=blue,citecolor=blue,urlcolor=blue]{hyperref}

\newcommand{\ket}[1]{|#1\rangle}
\newcommand{\bra}[1]{\langle #1|}
\newcommand{\ketbra}[2]{|#1\rangle\langle #2|}
\newcommand{\Tr}{\mathrm{Tr}}
\newcommand{\CREN}{\mathcal N^{\rm cr}}

\begin{document}

\title{Quantum Negativity Is Monogamous}

\author{Liuke Lyu}
\affiliation{Department of Physics, University of Toronto, Toronto, Ontario M5S 1A7, Canada}

\date{\today}

\begin{abstract}
He and Vidal conjectured that the squared negativity obeys $\mathcal N_{AB}^2+\mathcal N_{AC}^2\le\mathcal N_{A|BC}^2$ for every pure state of three parties. I prove this conjecture in all finite dimensions. This gives a computable monogamy relation in every dimension.
\end{abstract}

\maketitle

\noindent\textbf{Introduction}---Entanglement cannot be shared freely. If qubit $A$ is maximally entangled with $B$, it is uncorrelated with any third party $C$. Coffman, Kundu and Wootters (CKW) made this quantitative for three qubits: the squared concurrences obey $C_{AB}^2+C_{AC}^2\le C_{A|BC}^2$~\cite{CKW2000}, and Osborne and Verstraete extended the inequality to any number of qubits~\cite{OsborneVerstraete2006}. Beyond qubits, however, the concurrence is no longer monogamous. Three qutrits in the totally antisymmetric state $\psi=\sum_{abc}\epsilon_{abc}\ket{abc}/\sqrt6$ violate the CKW inequality~\cite{Ou2007}, and a qutrit $A$ with two qubits already violates it~\cite{KimDasSanders2009}. Squashed entanglement~\cite{ChristandlWinter2004} is monogamous in every dimension~\cite{KoashiWinter2004}, but it is NP-hard to compute~\cite{Huang2014}. Is there a computable entanglement measure that is monogamous in every dimension?

The negativity is a natural candidate~\cite{VidalWerner2002}. For a state $\rho$ of two parties $A$ and $B$, it is
\begin{equation*}
\mathcal N(\rho)=\frac{\|\rho^{T_A}\|_1-1}{2},
\end{equation*}
where $\rho^{T_A}$ is the partial transpose with respect to $A$, with matrix elements $\bra{i,k}\rho^{T_A}\ket{j,l}=\bra{j,k}\rho\ket{i,l}$ in a product basis, and $\|\cdot\|_1$ is the trace norm. Since $\Tr\rho^{T_A}=1$, $\mathcal N(\rho)$ is the sum of the absolute values of the negative eigenvalues of $\rho^{T_A}$. It is computable for any mixed state~\cite{VidalWerner2002}, and it vanishes for separable states~\cite{Peres1996,Horodecki1996}. For a pure tripartite state $\psi$, write $\mathcal N_{AB}=\mathcal N(\rho_{AB})$, $\mathcal N_{AC}=\mathcal N(\rho_{AC})$ and $\mathcal N_A=\mathcal N(\ketbra\psi\psi)$ across the cut $A|BC$. He and Vidal conjectured that
\begin{equation}
\mathcal N_{AB}^2+\mathcal N_{AC}^2\le\mathcal N_A^2
\label{eq:hv}
\end{equation}
for every pure state of any three parties~\cite{HeVidal2015} (rescaling $\mathcal N$, e.g.\ to $\|\rho^{T_A}\|_1-1$, leaves it unchanged); the same inequality, with negativities normalized by $d-1$, appeared earlier for qudits in Ref.~\cite{KimJooLee2011}. The antisymmetric state, which violates the CKW inequality, satisfies Eq.~\eqref{eq:hv} with room to spare: $\mathcal N_{AB}=\mathcal N_{AC}=1/3$ and $\mathcal N_A=1$~\cite{KimJooLee2011}. Ou and Fan proved Eq.~\eqref{eq:hv} for three qubits~\cite{OuFan2007}, and it is known for $2\otimes2\otimes2^n$~\cite{LuoLi2015} and $2\otimes2\otimes3$~\cite{GaoYanGao2021}. Numerical searches over $m\otimes m\otimes m$ states found no violation for $m\le4$~\cite{HeVidal2015}. Audenaert reformulated it as a block-matrix inequality and proved a special case with a qubit $A$ and $d_B=d_C$~\cite{Audenaert2015}. In general dimensions, Eq.~\eqref{eq:hv} has remained open~\cite{GuoJin2025}.

In this Letter, I prove Eq.~\eqref{eq:hv} for every pure state of three parties of any finite dimensions. The proof has three steps. First, an entanglement witness turns Eq.~\eqref{eq:hv} into the statement that a real symmetric matrix $M$, built from the Schmidt decomposition across $A|BC$, is copositive: $t^TMt\ge0$ for every vector $t$ with nonnegative entries. Second, I prove a property I call \emph{row dominance}: in every row of $M$, the diagonal entry outweighs the sum of any set of off-diagonal entries. It follows from one operator inequality that uses only $0\le P\le I$ for the witnesses. Finally, row dominance implies copositivity in one line.

\textbf{Main result}---I prove the following.

\begin{theorem}[Negativity monogamy]
\label{thm:hv}
For every pure state $\psi\in\mathbb C^{d_A}\otimes\mathbb C^{d_B}\otimes\mathbb C^{d_C}$, of any finite dimensions $d_A$, $d_B$, $d_C$, $\mathcal N_{AB}^2+\mathcal N_{AC}^2\le\mathcal N_A^2$.
\end{theorem}

\noindent The bound cannot be improved by a constant factor: $\psi=\Phi_{AB}\otimes\ket0_C$, with $\Phi_{AB}$ entangled, has $\mathcal N_{AC}=0$ and $\mathcal N_{AB}=\mathcal N_A$.

\textbf{Witness form}---Write $\psi=\sum_{i=0}^{r-1}s_i\ket i_A\ket{w_i}_{BC}$ in Schmidt form, with Schmidt rank $r\le\min(d_A,d_Bd_C)$, $s_i>0$ and both $\{\ket i_A\}$ and $\{\ket{w_i}_{BC}\}$ orthonormal. Then $\mathcal N_A=\sum_{i<j}s_is_j$~\cite{VidalWerner2002}, and
\begin{equation*}
\rho_{AB}^{T_A}=\sum_{ij}s_is_j\ketbra ji\otimes\Tr_C\ketbra{w_i}{w_j}.
\end{equation*}
Expand an operator $P$ on $AB$ in the Schmidt basis of $A$ as $P=\sum_{ij}\ketbra ij\otimes P_{ij}$, so that the block $P_{ij}=(\bra i_A\otimes I_B)P(\ket j_A\otimes I_B)$ is a $d_B\times d_B$ operator on $B$; throughout, $P_{ij}$ denotes such a block, not a matrix element of $P$. This gives $\Tr P\rho_{AB}^{T_A}=\sum_{ij}s_is_j\bra{w_j}P_{ij}\otimes I\ket{w_i}$. Likewise, an operator $Q$ on $AC$ gives $\Tr Q\rho_{AC}^{T_A}=\sum_{ij}s_is_j\bra{w_j}I\otimes Q_{ij}\ket{w_i}$. Now let $P$ and $Q$ be Hermitian operators with $0\le P,Q\le I$ (all eigenvalues in $[0,1]$), let $p,q\ge0$ with $p^2+q^2\le1$, and put $Z_{ij}=p\,P_{ij}\otimes I+q\,I\otimes Q_{ij}$, an operator on $BC$ with $Z_{ji}=Z_{ij}^\dagger$. Define the real symmetric $r\times r$ matrix
\begin{equation*}
M_{ii}=\bra{w_i}Z_{ii}\ket{w_i},\qquad M_{ij}=\tfrac12+\mathrm{Re}\bra{w_j}Z_{ij}\ket{w_i}\quad(i\ne j),
\end{equation*}
where $M_{ii}\ge0$ because $Z_{ii}\ge0$. Adding the two trace formulas above gives $p\Tr P\rho_{AB}^{T_A}+q\Tr Q\rho_{AC}^{T_A}=\sum_{ij}s_is_j\bra{w_j}Z_{ij}\ket{w_i}$, whose $(i,j)$ and $(j,i)$ terms are complex conjugates because $Z_{ji}=Z_{ij}^\dagger$. Together with $\mathcal N_A=\sum_{i<j}s_is_j$, which the $\frac12$ in $M_{ij}$ accounts for, this gives
\begin{equation}
s^TMs=\mathcal N_A+p\Tr P\rho_{AB}^{T_A}+q\Tr Q\rho_{AC}^{T_A}.
\label{eq:witness}
\end{equation}
Choose $P$ and $Q$ to be the projectors onto the negative eigenspaces of $\rho_{AB}^{T_A}$ and $\rho_{AC}^{T_A}$, so that the traces are $-\mathcal N_{AB}$ and $-\mathcal N_{AC}$, and let $(p,q)=(\mathcal N_{AB},\mathcal N_{AC})/(\mathcal N_{AB}^2+\mathcal N_{AC}^2)^{1/2}$ (if both vanish there is nothing to prove). Then $s^TMs=\mathcal N_A-(\mathcal N_{AB}^2+\mathcal N_{AC}^2)^{1/2}$. Since $s$ has positive entries, Theorem~\ref{thm:hv} follows once $M$ is copositive for all such $P$, $Q$, $p$ and $q$.

\textbf{Row dominance}---The key step is a bound on each row of $M$.

\begin{lemma}[Row dominance]
\label{lem:row}
For every row $i$ and every set $S$ of indices other than $i$, $M_{ii}+\sum_{j\in S}M_{ij}\ge0$.
\end{lemma}

\begin{proof}
\emph{(a) Block bounds.} The operator $P-P^2=P^{1/2}(I-P)P^{1/2}$ is positive semidefinite. In blocks,
\begin{equation*}
P-P^2=\sum_{ij}\ketbra ij\otimes\Big(P_{ij}-\sum_kP_{ik}P_{kj}\Big),
\end{equation*}
and each diagonal block is positive semidefinite: $P_{ll}-P_{ll}^2-\sum_{k\ne l}P_{lk}P_{kl}\ge0$ for every $l$. Each term $P_{lk}P_{kl}=P_{lk}P_{lk}^\dagger$ is positive semidefinite, and $0\le P_{ll}\le I$ gives $P_{ll}-P_{ll}^2\le P_{ll}$ and $P_{ll}-P_{ll}^2\le\frac14I$. Hence, for every $l$ and every $k\ne l$,
\begin{equation*}
\sum_{k'\ne l}P_{lk'}P_{k'l}\le P_{ll},\qquad P_{lk}P_{kl}\le\tfrac14I.
\end{equation*}
The same holds for $Q$.

\emph{(b) One column.} Fix $j\ne i$. Since $P_{ab}\otimes I$ commutes with $I\otimes Q_{cd}$,
\begin{equation*}
\begin{aligned}
Z_{ij}^\dagger Z_{ij}={}&p^2P_{ji}P_{ij}\otimes I+q^2I\otimes Q_{ji}Q_{ij}\\
&+pq\,(P_{ij}\otimes Q_{ji}+P_{ji}\otimes Q_{ij}).
\end{aligned}
\end{equation*}
With $R=P_{ji}\otimes I$ and $T=I\otimes Q_{ji}$, the cross term is $R^\dagger T+T^\dagger R\le R^\dagger R+T^\dagger T$, because $(R-T)^\dagger(R-T)\ge0$. By (a) and $p^2+q^2\le1$, this gives $\|Z_{ij}w_i\|^2\le\frac14+b_j$, where
\begin{equation*}
b_j=pq\bra{w_i}P_{ij}P_{ji}\otimes I+I\otimes Q_{ij}Q_{ji}\ket{w_i}\ge0.
\end{equation*}

\emph{(c) The whole row.} Summing over $j\ne i$ with (a), and using $pq\le p$ and $pq\le q$, gives $\sum_{j\ne i}b_j\le pq\bra{w_i}P_{ii}\otimes I+I\otimes Q_{ii}\ket{w_i}\le M_{ii}$. Since $\|w_j\|=1$, the Cauchy--Schwarz inequality and $\sqrt{1/4+b}\le\frac12+b$ for $b\ge0$ give
\begin{equation*}
|\bra{w_j}Z_{ij}\ket{w_i}|\le\|Z_{ij}w_i\|\le\tfrac12+b_j.
\end{equation*}
Hence $M_{ij}\ge-b_j$ for every $j\ne i$, and $M_{ii}+\sum_{j\in S}M_{ij}\ge M_{ii}-\sum_{j\ne i}b_j\ge0$.
\end{proof}

\noindent Lemma~\ref{lem:row} is sharp: equality holds for a nonempty $S$. For $\psi=(\ket{000}+x\ket{101}+y\ket{110})/\sqrt2$ with $x^2+y^2=1$, take $P$ and $Q$ to be the projectors onto the singlet $(\ket{01}-\ket{10})/\sqrt2$ of $AB$ and of $AC$, and $(p,q)=(y,x)$. Here $w_0=\ket{00}$ and $w_1=x\ket{01}+y\ket{10}$. Both projectors vanish on $\ket0_A\ket{w_0}$, so $M_{00}=0$, while $\bra{w_1}Z_{01}\ket{w_0}=-\frac12$, so $M_{01}=\frac12-\frac12=0$. Hence $M_{00}+M_{01}=0$ with $S=\{1\}$.

\textbf{Copositivity}---I can now finish the proof.

\begin{proof}[Proof of Theorem~\ref{thm:hv}]
Let $t\in\mathbb R^r$ have nonnegative entries, and let $S_i$ be the set of $j\ne i$ with $M_{ij}<0$. Dropping the nonnegative off-diagonal terms, and using $t_it_j\le(t_i^2+t_j^2)/2$ on the negative ones,
\begin{equation*}
\begin{aligned}
t^TMt&\ge\sum_it_i^2M_{ii}+\sum_{i\ne j}\frac{t_i^2+t_j^2}{2}\min(M_{ij},0)\\
&=\sum_it_i^2\Big(M_{ii}+\sum_{j\in S_i}M_{ij}\Big)\ge0,
\end{aligned}
\end{equation*}
where the equality uses $M_{ij}=M_{ji}$ and the last step is Lemma~\ref{lem:row}. So $M$ is copositive, and Eq.~\eqref{eq:witness} at $t=s$ gives $\mathcal N_A\ge(\mathcal N_{AB}^2+\mathcal N_{AC}^2)^{1/2}$.
\end{proof}

\textbf{Conclusion}---I have proved the He--Vidal monogamy of the squared negativity for every pure state of three parties of any finite dimensions. The key step is to fix the witnesses first: the inequality then becomes the copositivity of a matrix in the Schmidt coefficients of $A$, and every row of this matrix is dominated by its diagonal entry, whatever the witnesses and the dimensions. The negativity thus gives a monogamy relation that holds in every dimension, where the concurrence fails.

Two consequences follow at once. First, the residual $\pi_A=\mathcal N_A^2-\mathcal N_{AB}^2-\mathcal N_{AC}^2$ is nonnegative in all dimensions, and so is the negativity-based three-tangle of Ou and Fan, the average of $\pi_A$, $\pi_B$ and $\pi_C$~\cite{OuFan2007}, previously established only in the special cases above. Second, the theorem extends to mixed states. Let $\rho=\sum_kp_k\ketbra{\psi_k}{\psi_k}$ be any decomposition of a state $\rho$ of $ABC$ into pure states. Since the negativity is convex, the triangle inequality in $\mathbb R^2$ and Theorem~\ref{thm:hv} give
\begin{equation*}
\begin{aligned}
\big(\mathcal N_{AB}^2+\mathcal N_{AC}^2\big)^{1/2}&\le\sum_kp_k\big(\mathcal N_{AB}(\psi_k)^2+\mathcal N_{AC}(\psi_k)^2\big)^{1/2}\\
&\le\sum_kp_k\mathcal N_A(\psi_k).
\end{aligned}
\end{equation*}
This holds for every decomposition, and the smallest right side is the convex-roof extended negativity $\CREN_{A|BC}(\rho)=\min\sum_kp_k\mathcal N_A(\psi_k)$~\cite{LeeChiOhKim2003}. Hence
\begin{equation}
\mathcal N_{AB}^2+\mathcal N_{AC}^2\le\CREN_{A|BC}(\rho)^2.
\label{eq:mixed}
\end{equation}
Some questions remain open. My argument does not extend directly to more parties. The states that attain equality in Eq.~\eqref{eq:hv} are also not characterized.

\begin{acknowledgments}
The main theorem and its proof were developed in collaboration with Claude Opus~5.5 (Anthropic) and Gemini (Google); the copositive form of Eq.~\eqref{eq:witness} was found by Gemini, and the row-dominance argument of Lemma~\ref{lem:row} was found with Claude. L.L.\ acknowledges support from the University of Toronto Arts \& Science Postdoctoral Fellowship Program. Claude (Anthropic) was also used to assist with the numerical checks, and both Claude (Anthropic) and Gemini (Google) were used to assist in polishing and improving the readability of the manuscript. The generated code was reviewed, tested, and validated by the author, who takes full responsibility for the software, analyses, integrity, and conclusions reported in this work.
\end{acknowledgments}

\bibliographystyle{apsrev4-2}
\bibliography{references}

\end{document}